\documentclass[aps,prd,12point,nofootinbib,superscriptaddress]{revtex4-2}

\usepackage[dvipsnames]{xcolor}
\usepackage{graphicx}
\usepackage{amssymb,bm}
\usepackage{amsmath,amsthm}
\usepackage{epstopdf}
\usepackage{hyperref}
\hypersetup{
    colorlinks=true,
    linkcolor=blue,
    filecolor=magenta,
    urlcolor=cyan,
    citecolor=magenta
}
\newcommand{\be}{\begin{equation}} 
\newcommand{\ee}{\end{equation}}

\begin{document}
\def\theequation{\thesection.\arabic{equation}}
\def\theequation{\arabic{section}.\arabic{equation}}

\title{Past-directed scalar field gradients and 
scalar-tensor thermodynamics}

\author{Andrea Giusti}
\email[]{agiusti@phys.ethz.ch}
\affiliation{Institute for Theoretical Physics, ETH Zurich,
Wolfgang-Pauli-Strasse 27, 8093, Zurich, Switzerland}

\author{Serena Giardino}
\email[]{serena.giardino@aei.mpg.de}
\affiliation{Max Planck Institute for Gravitational Physics (Albert Einstein 
Institute), Callinstra{\ss}e 38, 30167 Hannover, Germany}

\author{Valerio Faraoni}
\email[]{vfaraoni@ubishops.ca}
\affiliation{Department of Physics \& Astronomy, Bishop's University, 
2600 College Street, Sherbrooke, Qu\'ebec, Canada J1M~1Z7 }

\begin{abstract}

We refine and slightly enlarge the recently proposed first-order 
thermodynamics of scalar-tensor gravity to include gravitational scalar 
fields with timelike and past-directed gradients. The implications and 
subtleties arising in this situation are discussed and an exact 
cosmological solution of scalar-tensor theory in first-order 
thermodynamics is revisited in light of these results.

\end{abstract}

\maketitle

\section{Introduction}
\label{sec:1}
\setcounter{equation}{0}

There are several motivations to extend the theory of gravity beyond 
Einstein's General Relativity (GR). All 
attempts to 
reconcile  this theory with  quantum physics 
introduce deviations from GR in the form of extra 
fields, higher-order equations of motion, or higher-order curvature 
invariants. For example, taking the low-energy limit of the 
bosonic string theory, the simplest among string theories, produces 
 $\omega =-1$ Brans-Dicke theory instead of GR, which is the 
prototype of a scalar-tensor theory (and $\omega$ is the Brans-Dicke 
coupling) \cite{Callan:1985ia, Fradkin:1985ys}.

However, the most compelling motivations to study alternative 
theories of gravity come from cosmology. For instance, the inflationary 
model most favoured by data, namely Starobinsky inflation, includes 
quantum corrections to GR. Most importantly, a satisfactory understanding 
of the present-day accelerated expansion of the universe is lacking within 
the realm of the standard $\Lambda$CDM model of cosmology based on GR: it 
requires one to introduce an astonishingly fine-tuned cosmological 
constant or another form of {\em ad hoc} dark energy, whose nature remains 
elusive \cite{AmendolaTsujikawabook}.

In any case, even admitting the presence of dark energy still leaves other 
problems of $\Lambda$CDM unresolved, such as the Hubble tension 
\cite{Riess:2019qba,DiValentino:2021izs}, the requirement for an equally 
mysterious dark matter, and the singularity problem that plagues cosmology 
and black hole physics. It is at least reasonable, therefore, to study 
alternative theories of gravity to resolve or alleviate these issues.

The simplest way to modify GR consists of adding a scalar (massive) degree 
of freedom, which resulted in Brans-Dicke gravity \cite{Brans:1961sx} and 
its scalar-tensor generalizations \cite{Bergmann:1968ve,Nordtvedt:1968qs,Wagoner:1970vr,Nordtvedt:1970uv}. The class of $f(R)$ 
theories of gravity, which turns out to be a subclass of scalar-tensor 
theories, is extremely popular to explain the present cosmic acceleration 
without dark energy (\cite{Capozziello:2003tk}, see 
\cite{Sotiriou:2008rp,DeFelice:2010aj,Nojiri:2010wj} for reviews). In the 
last decade, the old Horndeski gravity \cite{Horndeski} has been 
revisited 
and studied intensively (see \cite{Kobayashi:2019hrl} for a review). This 
class of theories was 
believed to be the most general scalar-tensor gravity admitting 
second-order equations of motion, but then it was discovered that, if a 
suitable degeneracy condition is satisfied, the even more general 
Degenerate Higher Order Scalar-Tensor (DHOST) theories admit second-order 
equations of motion (see \cite{DHOSTreview1} for  a review).

Horndeski and DHOST theories contain arbitrary functions in their actions 
that make the field equations very cumbersome and their study difficult. 
The multi-messenger event GW170817/GRB170817, 
\cite{LIGOScientific:2017vwq,LIGOScientific:2017zic} 
 confirming that 
gravitational wave modes propagate at the speed of light, has essentially 
ruled out Horndeski theories with the most 
complicated structure \cite{Langlois:2017dyl}, but many possibilities 
(corresponding to four free functions in the action) remain. As a result, 
it is difficult to grasp the detailed physical meaning of these theories 
and their solutions and much of the work necessarily remains confined 
to formal theoretical aspects and to the search for analytical 
solutions.

In an attempt to  gain physical intuition for scalar-tensor gravity  
(including viable Horndeski
theories), it is 
fruitful to  interpret it  through an effective fluid description in which 
the (Jordan frame) field 
equations are written as effective Einstein equations and the remaining 
geometrical terms, when moved to the right-hand side, form the 
stress-energy tensor of an effective {\em dissipative} fluid 
\cite{Pimentel:1989bm,Faraoni:2018qdr,Nucamendi:2019uen,Giusti:2021sku}. 
This effective fluid description is possible when the gradient of the 
scalar field degree of freedom $\phi$ of the theory is timelike 
\cite{Pimentel:1989bm,Faraoni:2018qdr,Nucamendi:2019uen}.

In 
this context, using the three constitutive relations postulated in   
Eckart's first-order thermodynamics of dissipative 
fluids \cite{Eckart:1940te}, we were able to introduce an effective 
``temperature of gravity'' and shear and bulk viscosity coefficients 
\cite{Faraoni:2021lfc,Faraoni:2021jri,Giusti:2021sku}.  Armed with these 
concepts, one can describe GR as the state of 
thermal equilibrium of gravity corresponding to zero temperature and 
scalar-tensor gravity as a state of non-equilibrium at higher temperature 
(this temperature being relative to GR) 
\cite{Faraoni:2021lfc,Faraoni:2021jri,Giusti:2021sku}.\footnote{The idea 
of modified (specifically, quadratic $f(R)$) gravity as a state of 
non-equilibrium, with GR as the equilibrium state,  
goes back to Jacobson's thermodynamics of spacetime 
\cite{Jacobson:1995ab,Eling:2006aw}. However, in 
spite of many studies, a ``temperature of gravity'' was never identified 
in this context, which is entirely different from the first-order 
thermodynamics of scalar-tensor gravity discussed here.}  Dissipation 
corresponds  to the relaxation of this effective fluid toward the GR state 
of equilibrium.

We applied the recent first-order thermodynamics of modified gravity to 
Friedmann-Lema\^itre-Robertson-Walker (FLRW) cosmology 
\cite{Giardino:2022sdv} and additionally searched for possible alternative 
equilibrium states, which turned out to be metastable 
\cite{Faraoni:2022jyd,Faraoni:2022doe}. We also found that the 
alternative description of scalar-tensor gravity in the Einstein conformal 
frame swaps temperature with chemical potential \cite{Faraoni:2022gry}.

These early studies do not adequately discuss a potential limitation of 
the formalism, {\em i.e.}, the fact that the scalar field gradient must be 
future-directed for it to be meaningful.

Here we refine the previous discussions to make this limitation explicit 
and we discuss a possible extension of the formalism to situations in 
which this gradient is timelike but past-directed, which does indeed occur 
in certain analytical solutions of scalar-tensor theories.  We follow the 
notations of Ref.~\cite{Waldbook}: the metric signature is ${-}{+}{+}{+}$ 
and we use units in which the speed of light $c$ and Newton's constant $G$ 
are unity.

\section{Past-directed scalar field gradients}
\label{sec:2}
\setcounter{equation}{0}

We consider scalar-tensor theories described by the action
\be
S_{\rm ST} = \frac{1}{16\pi} \int d^4x \sqrt{-g} \left[ \phi R 
-\frac{\omega(\phi )}{\phi} 
\, \nabla^c\phi \nabla_c\phi -V(\phi) \right]+S^{\rm (m)} \,, \label{STaction}
\ee
where $R$ is the Ricci scalar, $V(\phi)$ the scalar field potential and 
$S^\text{(m)}=\int d^4x \sqrt{-g} \,  {\cal L}^\text{(m)} $ the matter 
action.  Let us define the timelike vector field $u^a$ as
\begin{equation}
u^a := \frac{\nabla ^a \phi}{\sqrt{2  X}} \, , \qquad X := - \frac{1}{2} 
\, \nabla_a \phi \nabla^a \phi >0  \,.
\end{equation}
Assuming that the spacetime manifold $(\mathcal{M}, g_{ab})$ admits a 
chart $(t, 
\bm{x})$ with time coordinate $t$, then $g_{ab} \, u^a \, (\partial _t)^b 
> 0$ 
implies that $u^a$ is past-directed and it cannot be identified with the 
4-velocity of an effective fluid, which is defined as timelike 
{\em future-directed} vector field. 

Now, let the scalar field $\phi$ be such that 
$\nabla^a \phi$ is past-directed: we can then define a future-directed 
vector field as
\begin{equation}
v^a := - u^a = - \frac{\nabla ^a \phi}{\sqrt{2  X}} \, .
\end{equation}
The corresponding projection operator onto the 3-space orthogonal to $v^a$ 
is ${\mathfrak{h}^a}_b$, where 
\begin{equation}
\mathfrak{h}_{ab} := g_{ab} + v_a v_b = g_{ab} + u_a u_b = g_{ab} + 
\frac{\nabla _a \phi \nabla _b 
\phi}{2 X}
=h_{ab} \, ,
\end{equation}
and ${h^a}_b$ is the projection operator onto the 3-space orthogonal to 
$u^a$. Thus, $h_{ab}$ remains unaffected by the change of sign in the 
definition of the 4-velocity when the timelike gradient $\nabla^a \phi$ is 
past-directed instead of being future-directed.
\\

\subsection{Kinematic quantities}

Let us examine now how the kinematic quantities associated with the 
effective scalar-tensor dissipative fluid 
\cite{Pimentel:1989bm,Faraoni:2018qdr,Nucamendi:2019uen} change when the 
definition of 4-velocity is modified to account for a past-directed 
gradient $\nabla^a\phi$. In particular, we make explicit the relations 
between the kinematic quantities associated with $v^a$ 
(denoting them with $^{(v)}$) and those 
corresponding to $u^a=-v^a$ (denoting them with $^{(u)}$). 
For the 4-velocity gradient, we have 

\begin{equation}
\nabla_a v_b = - \nabla_a u_b = - \frac{1}{\sqrt{2 X}} \left( \nabla _a 
\nabla _b \phi - 
\frac{\nabla_a X \nabla _b \phi}{2 X} \right) \, ,
\end{equation} 
which implies
\begin{equation}
\Theta_{(v)} = \nabla _a v^a = - \nabla _a u^a = - \Theta_{(u)} 
\end{equation}
for the expansion scalar of the effective fluid, 
\begin{equation}
a_{(v)} ^a := v^c \nabla_c v^a = u^c \nabla_c u^a = a_{(u)} ^a 
\end{equation}
for its 4-acceleration, while the projection of the velocity gradient onto 
the 3-space of the comoving observers reads 
\begin{equation}
V^{(v)}_{ab} := {h_a}^c {h_b}^d \nabla _d v_c = - {h_a}^c {h_b}^d \nabla 
_d u_c = - V^{(u)}_{ab} \, ,
\end{equation}
and the new shear tensor is  
\begin{equation}
\sigma ^{(v)} _{ab} := V ^{(v)}_{(ab)} - \frac{\Theta^{(v)} }{3} \, 
h_{ab} = 
- \left( V^{(u)} _{ab} - \frac{\Theta^{(u)} }{3} \, 
h_{ab} \right) = - \sigma ^{(u)} _{ab} \, .
\end{equation}
These kinematic quantities do not depend on the field equations and are the 
same in all scalar-tensor gravity theories.

\subsection{Effective fluid stress-energy tensor}

The effective energy-momentum tensor for scalar-tensor gravity reads
\begin{equation}
8\pi T_{ab} = \frac{\omega}{\phi^2} \left( \nabla_a \phi 
\nabla_b \phi -
 \frac{1}{2} \, g_{ab} \nabla^c \phi \nabla_c \phi \right)  + 
 \frac{1}{\phi} \left( \nabla_a \nabla_b \phi -g_{ab} \square \phi \right)
- \frac{V}{2 \phi} \, g_{ab}
\end{equation}
and it has been recognised to have the form of an imperfect fluid 
stress-energy tensor. In the case of past-directed gradients of $\phi$, 
we can write it as 
\be 
T^{(v)}_{ab} = \rho^{(v)} \, v_a v_b + q^{(v)}_a v_b 
+ q^{(v)}_b v_a + \Pi^{(v)}_{ab} 
\,,\label{imperfectTab} 
\ee 
where the effective energy density, heat flux density, stress tensor, 
isotropic pressure, and anisotropic stress tensor (the trace-free part 
$\pi_{ab}$ of the stress tensor $\Pi_{ab}$) in the comoving frame of 
the effective fluid are, respectively,  
\begin{eqnarray} 
\rho ^{(v)} &=& T_{ab}    v^a v^b \,, \nonumber \label{rhophi}\\ 
&&\nonumber\\
q^{(v)} _a & =& -T_{cd} \, v^c 
{h_a}^d \,, \nonumber 
  \label{qphi}\\
&&\nonumber\\
 \Pi ^{(v)}_{ab} &= & P^{(v)} h_{ab} + \pi ^{(v)}_{ab} = T_{cd} \, 
{h_a}^c \, {h_b}^d \,,  \nonumber
\label{Piphi}\\
&&\nonumber\\
    P^{(v)} &=& \frac{1}{3}\, g^{ab}\Pi ^{(v)}_{ab} =\frac{1}{3} \, 
h^{ab} T_{ab} \,, \nonumber
\label{Pphi}\\
&&\nonumber\\
    \pi ^{(v)}_{ab} &=& \Pi ^{(v)} _{ab} - P^{(v)}h_{ab} \,. 
\label{piphi} \nonumber
\end{eqnarray} 

It is straightforward to see that some of these quantities are not 
altered with respect to those arising from future-directed scalar field 
gradients:
\be
\label{relatio}
\rho^{(v)}  = \rho ^{(u)} \,, \quad\quad\quad \Pi^{(v)}_{ab} = \Pi 
^{(u)}_{ab} \,,
\quad\quad\quad P^{(v)} = P^{(u)} \,, \quad\quad\quad \pi^{(v)}_{ab} = \pi 
^{(u)}_{ab} \,.
\ee
However, the heat flux density changes sign when the 4-velocity changes 
orientation:
\be
\label{qs}
q^{(v)} _a = -T_{cd} \, v^c {h_a}^d = T_{cd} \, u^c {h_a}^d = - q^{(u)} 
_a \, ,
\ee
which has important consequences for the definition of a meaningful 
temperature, as we detail in the following.

\subsection{First-order thermodynamics with past-directed scalar field 
gradients}

Eckart's thermodynamics provides a simple non-equilibrium thermodynamics 
that is first-order in the dissipative variables. Based on its three 
constitutive equations \cite{Eckart:1940te}, we can find relationships 
between the kinematic quantities of the effective fluid and the 
dissipative thermodynamical variables, therefore building an analogy 
between the imperfect fluid and scalar-tensor gravity in a thermodynamical 
setting.  The following relationships between the 
heat flux and the 4-acceleration, and the anisotropic stress and shear 
tensor, respectively, hold when dealing with
future-directed gradients 
\cite{Faraoni:2021lfc,Faraoni:2021jri,Giusti:2021sku,Giardino:2022sdv}:
\be
q_a^{(u)} = -\frac{ \sqrt{2X}}{ 8 \pi \phi} \, a^{(u)}_a 
\,, \quad\quad\quad 
\pi ^{(u)}_{ab} = \frac{ \sqrt{2X}}{ 8 \pi \phi} \, \sigma ^{(u)} _{ab} \,.
\ee
Hence, given \eqref{relatio} and \eqref{qs}, we have 
\be
q^{(v)} _a = - q^{(u)} _a = 
\frac{ \sqrt{2X}}{ 8 \pi \phi} \, a^{(u)}_a 
= \frac{ \sqrt{2X}}{ 8 \pi \phi} \, a^{(v)}_a 
\ee
and
\be
\pi ^{(v)}_{ab} = \pi ^{(u)}_{ab} = \frac{ \sqrt{2X}}{ 8 \pi \phi} \, 
\sigma ^{(u)} _{ab} = - \frac{ \sqrt{2X}}{ 8 \pi \phi} \, \sigma ^{(v)} 
_{ab} \, .
\ee
This means that, {\em for a scalar field with timelike past-directed 
gradient}, one finds the ``temperature of scalar-tensor 
gravity''
\be
(\mathcal{K} \mathcal{T}) ^{(v)} = - (\mathcal{K} \mathcal{T}) ^{(u)} = - 
\frac{ \sqrt{2X}}{ 8 \pi \phi} < 0 
\ee
(where ${\cal K}$ is the thermal conductivity, ${\cal T}$ is the 
temperature, and these two quantities always appear together in our 
analysis).  Moreover, the shear viscosity coefficient $\eta$  
defined by the constitutive relation $\pi_{ab} = -2\eta \, \sigma_{ab}$  
\cite{Eckart:1940te} reads 
\be 
\eta^{(v)} = - \eta^{(u)} = \frac{ \sqrt{2X}}{ 16 \pi \phi} > 0 \, .
\ee
Thus, for past-directed gradients, we find a negative temperature 
and positive shear viscosity, opposite to the result for future-directed 
gradients. This is precisely the reason why making sure the velocity of 
$\phi$ is future-directed is crucial: the thermodynamical analogy built 
in \cite{Faraoni:2021lfc,Faraoni:2021jri,Giusti:2021sku} itself relies on 
a meaningful notion of temperature. The fact that such a temperature 
naturally arose to be positive-definite in the case of future-directed 
velocity is one of the promising features of the formalism. Moreover, 
modified gravity theories with degrees of freedom additional to those of 
GR always have a positive temperature with respect to GR, which is quite 
intuitive. The only meaningful situation where we found a negative 
temperature was that of Nordstr\"{o}m gravity, that possesses 
\textit{less} degrees of freedom than GR \cite{Faraoni:2022doe}. 
Additionally, the negative shear viscosity previously found in 
\cite{Faraoni:2021jri,Giusti:2021sku} made sense as there is no reason to 
expect the effective fluid we are dealing with to be isolated. In the 
present work, a positive $\eta$ corresponds to the more usual case of an 
isolated fluid.

The previous findings 
\cite{Faraoni:2021lfc,Faraoni:2021jri,Giusti:2021sku,Giardino:2022sdv} 
remain valid, provided that one restricts the application of our formalism to scalar fields with future-directed timelike gradients.
\\

\subsection{Approach to thermal equilibrium}

Let us compute now the evolution of $(\mathcal{K} \mathcal{T})^{(v)}$ 
with  respect to the time direction dictated by $v^a$, which is given by 
\cite{Faraoni:2021lfc,Faraoni:2021jri,Giusti:2021sku}
\begin{equation*}
\begin{split}
\frac{d}{d\tau} (\mathcal{K} \mathcal{T}) ^{(v)} &= v^a \nabla _a 
(\mathcal{K} \mathcal{T}) ^{(v)} = (- u^a) \nabla _a \left[ - 
(\mathcal{K} \mathcal{T}) ^{(u)} \right] 
= u^a \nabla _a (\mathcal{K} \mathcal{T}) ^{(u)} \\
&= 8\pi(\mathcal{K} \mathcal{T}) _{(u)} ^2 - \Theta _{(u)} (\mathcal{K} 
\mathcal{T}) _{(u)} + \frac{\Box \phi}{8 \pi \phi} \\
&= 8\pi(\mathcal{K} \mathcal{T}) _{(v)} ^2 - \Theta _{(v)} (\mathcal{K} 
\mathcal{T}) _{(v)} + \frac{\Box \phi}{8 \pi \phi} \, .
\end{split}
\end{equation*}
That is, the effective heat equation describing the approach to 
(or the departure from) 
thermal equilibrium in the first-order thermodynamics of 
scalar-tensor gravity reads 
\be
\frac{d}{d\tau} (\mathcal{K} \mathcal{T}) ^{(v)} = 8\pi(\mathcal{K} 
\mathcal{T}) _{(v)} ^2 - \Theta _{(v)} (\mathcal{K} \mathcal{T}) _{(v)} + 
\frac{\Box \phi}{8 \pi \phi} 
\ee
and is, therefore, not 
affected by the replacement $u^a \mapsto -u^a =v^a $.

\section{Revisiting the Brans-Dicke dust solution}
\label{sec:4}
\setcounter{equation}{0}

Many analytical solutions of scalar-tensor gravity are known in various 
physical contexts, ranging from FLRW cosmology ({\em e.g.}, \cite{mybook1}) 
to spherically symmetric solutions describing black holes and other 
objects (see \cite{Faraoni:2021nhi} for a recent review). A large fraction 
of the literature on Horndeski and DHOST gravity is devoted to the search 
of such solutions with disformal (and other) techniques 
\cite{BenAchour:2020wiw, Faraoni:2021gdl, Achour:2021pla, Babichev:2012re,Anabalon:2013oea, Babichev:2013cya, Charmousis:2014zaa, Kobayashi:2014eva,Babichev:2016kdt, Motohashi:2018wdq, Babichev:2017guv, Anson:2020trg, BenAchour:2020fgy,Chatzifotis:2021hpg,Saadati:2022jdc}. These solutions 
were of course derived regardless of the effective fluid 
formalism and  the 
first-order thermodynamics approach which, {\em per se}, do not offer 
new methods for solving the field equations exactly (although they do 
offer novel physical 
interpretations \cite{Miranda:2022wkz}). 

Therefore, certain scalar-tensor solutions feature timelike and 
past-directed gradient $\nabla^a \phi$ of the scalar 
degree of freedom. Systematically searching for these solutions and 
listing them would not be 
particularly illuminating; rather, we 
focus on a classic simple FLRW solution that has been known for a long 
time, namely the Brans-Dicke dust solution \cite{Brans:1961sx}, 
which we analysed in \cite{Giardino:2022sdv}. This is the only 
solution we studied through the lens of first-order thermodynamics whose study 
requires to be revisited in light of the extension to past-directed fluid 
velocity that we provide in this paper.

This solution describes a matter-dominated universe permeated by a pressureless dust fluid in Brans-Dicke gravity 
\cite{Brans:1961sx} with $V(\phi)=0$ and $\omega\neq -4/3$ and reads
\be
a(t)=a_0 \, t^q\,, \quad\quad\quad \phi(t)=\phi_0 \, t^s \,, 
\quad\quad\quad \rho^\mathrm{(m)}(t)= \rho_0 \, t^r \,,
\ee
where $a(t)$ is the cosmic scale factor, $\rho^{(\rm 
m)}=\rho^{(\rm m)}(t)$ the matter energy density, $\rho_0= C/a_0^3 $, 
$C$ is an integration constant related to the  initial conditions, and
\be
q=\dfrac{2(\omega+1)}{3\omega+4} \,, \quad\quad\quad 
s=\dfrac{2}{3\omega+4} \,, \quad\quad\quad r=-3q
\ee
satisfy $3q+s=2$. Then, if the dot denotes
differentiation with respect to $t$, we find
\be
\dot{\phi} = \frac{s}{t} \, \phi \,.  
\ee
Since $\phi >0$, in order to have $\dot{\phi}>0$ one has to require  
$s>0$, which implies $\omega>-4/3$. From these assumptions, it 
follows that 
\be
\nabla^a \phi = g^{a0} \dot{\phi} = g^{a0} \, \frac{s \phi}{t}  \, ,
\ee
which implies $\nabla^0 \phi = - s\phi/t <0$. Therefore $\nabla^a \phi$ is 
past-directed.

The 4-velocity of the effective fluid must therefore be defined as  
\be
v^a = - u^a = - \frac{\nabla^a \phi}{\sqrt{2X}} \, , \quad 2 X := - 
\nabla^a \phi \nabla_a \phi 
= \frac{s^2 \phi^2}{t^2} \, .
\ee
The product of the temperature and the thermal conductivity 
is therefore negative and diverges as the  initial cosmological 
singularity is approached 
\be
(\mathcal{K} \mathcal{T}) ^{(v)} = - \frac{ \sqrt{2X}}{ 8 \pi \phi} = - 
\frac{s}{8 \pi \, t} \underset{t \to 0^+}{\longrightarrow} - \infty \, .
\ee
The shear viscosity $\eta^{(v)}$ vanishes because of the symmetries of FLRW (the heat flux $q^{(v)}$ would vanish too, 
but in \cite{Giardino:2022sdv}  we chose the heat flux as a timelike 
vector aligned with the four-velocity of
comoving observers).

\section{Conclusions}
\label{sec:5}

In this work we extend the first-order thermodynamics of 
scalar-tensor gravity to timelike and past-directed gradients of the 
scalar field. The first-order thermodynamics is a recent approach that 
aims to provide a novel understanding of the intriguing relationship 
between thermodynamics and gravity, by characterising GR as a 
zero-temperature equilibrium state and any modified gravity with 
additional degrees of freedom than GR as a non-equilibrium state. This 
idea depicts a "thermodynamics of gravitational theories" and is based on 
the imperfect fluid description of scalar-tensor theories.

Since the whole picture relies on the analogy of modified theories with an 
effective scalar field fluid that is supposed to have a meaningful 
(\textit{i.e.} future-directed) 4-velocity, previous works had not 
considered the possibility of past-directed velocity. However, since past-directed gradients do arise in some exact solutions of scalar-tensor gravity, it is worth trying to encompass them into an extension of our formalism, while clearly delineating its boundaries of 
applicability.

The present work fills this gap: we find that the kinematic fluid 
quantities remain unchanged, but some thermodynamical variables like heat 
fluxes change sign, leading to a negative temperature and a positive shear 
viscosity, at variance with previous works. These results are also 
confirmed by revisiting the Brans-Dicke dust solution, which does have a 
past-directed fluid velocity.

A negative temperature is problematic in our 
formalism where additional degrees of freedom to those of GR give modified 
theories a positive-definite temperature. We cannot provide an assessment of the physical viability of solutions in scalar-tensor gravity through the sign of the temperature within our formalism, but we are now aware of the need to restrict upcoming analyses to situations with future-directed scalar field velocity 
only, if the thermodynamical formalism is to meaningfully hold.

\setcounter{equation}{0}


\section*{Acknowledgments}

A.G.~is supported by the European Union's Horizon 2020 research and 
innovation programme under the Marie Sk\l{}odowska-Curie Actions (grant 
agreement No.~895648--CosmoDEC). The work of A.G~has also been carried out 
in the framework of the activities of the Italian National Group of 
Mathematical Physics [Gruppo Nazionale per la Fisica Matematica (GNFM), 
Istituto Nazionale di Alta Matematica (INdAM)]. S.G. thanks Jean-Luc Lehners at
AEI Potsdam for hospitality. V.F. is supported by the Natural 
Sciences \& Engineering Research Council of Canada (grant 2016-03803).

\bibliographystyle{apsrev4-1}
\bibliography{refe}{}

\begin{thebibliography}{52}%
\makeatletter
\providecommand \@ifxundefined [1]{%
 \@ifx{#1\undefined}
}%
\providecommand \@ifnum [1]{%
 \ifnum #1\expandafter \@firstoftwo
 \else \expandafter \@secondoftwo
 \fi
}%
\providecommand \@ifx [1]{%
 \ifx #1\expandafter \@firstoftwo
 \else \expandafter \@secondoftwo
 \fi
}%
\providecommand \natexlab [1]{#1}%
\providecommand \enquote  [1]{``#1''}%
\providecommand \bibnamefont  [1]{#1}%
\providecommand \bibfnamefont [1]{#1}%
\providecommand \citenamefont [1]{#1}%
\providecommand \href@noop [0]{\@secondoftwo}%
\providecommand \href [0]{\begingroup \@sanitize@url \@href}%
\providecommand \@href[1]{\@@startlink{#1}\@@href}%
\providecommand \@@href[1]{\endgroup#1\@@endlink}%
\providecommand \@sanitize@url [0]{\catcode `\\12\catcode `\$12\catcode
  `\&12\catcode `\#12\catcode `\^12\catcode `\_12\catcode `\%12\relax}%
\providecommand \@@startlink[1]{}%
\providecommand \@@endlink[0]{}%
\providecommand \url  [0]{\begingroup\@sanitize@url \@url }%
\providecommand \@url [1]{\endgroup\@href {#1}{\urlprefix }}%
\providecommand \urlprefix  [0]{URL }%
\providecommand \Eprint [0]{\href }%
\providecommand \doibase [0]{http://dx.doi.org/}%
\providecommand \selectlanguage [0]{\@gobble}%
\providecommand \bibinfo  [0]{\@secondoftwo}%
\providecommand \bibfield  [0]{\@secondoftwo}%
\providecommand \translation [1]{[#1]}%
\providecommand \BibitemOpen [0]{}%
\providecommand \bibitemStop [0]{}%
\providecommand \bibitemNoStop [0]{.\EOS\space}%
\providecommand \EOS [0]{\spacefactor3000\relax}%
\providecommand \BibitemShut  [1]{\csname bibitem#1\endcsname}%
\let\auto@bib@innerbib\@empty
\bibitem [{\citenamefont {Callan}\ \emph {et~al.}(1985)\citenamefont {Callan},
  \citenamefont {Martinec}, \citenamefont {Perry},\ and\ \citenamefont
  {Friedan}}]{Callan:1985ia}%
  \BibitemOpen
  \bibfield  {author} {\bibinfo {author} {\bibfnamefont {C.~G.}\ \bibnamefont
  {Callan}, \bibfnamefont {Jr.}}, \bibinfo {author} {\bibfnamefont {E.~J.}\
  \bibnamefont {Martinec}}, \bibinfo {author} {\bibfnamefont {M.~J.}\
  \bibnamefont {Perry}}, \ and\ \bibinfo {author} {\bibfnamefont
  {D.}~\bibnamefont {Friedan}},\ }\href {\doibase 10.1016/0550-3213(85)90506-1}
  {\bibfield  {journal} {\bibinfo  {journal} {Nucl. Phys. B}\ }\textbf
  {\bibinfo {volume} {262}},\ \bibinfo {pages} {593} (\bibinfo {year}
  {1985})}\BibitemShut {NoStop}%
\bibitem [{\citenamefont {Fradkin}\ and\ \citenamefont
  {Tseytlin}(1985)}]{Fradkin:1985ys}%
  \BibitemOpen
  \bibfield  {author} {\bibinfo {author} {\bibfnamefont {E.~S.}\ \bibnamefont
  {Fradkin}}\ and\ \bibinfo {author} {\bibfnamefont {A.~A.}\ \bibnamefont
  {Tseytlin}},\ }\href {\doibase 10.1016/0550-3213(85)90559-0} {\bibfield
  {journal} {\bibinfo  {journal} {Nucl. Phys. B}\ }\textbf {\bibinfo {volume}
  {261}},\ \bibinfo {pages} {1} (\bibinfo {year} {1985})},\ \bibinfo {note}
  {[Erratum: Nucl.Phys.B 269, 745--745 (1986)]}\BibitemShut {NoStop}%
\bibitem [{\citenamefont {Amendola}\ and\ \citenamefont
  {Tsujikawa}(2010)}]{AmendolaTsujikawabook}%
  \BibitemOpen
  \bibfield  {author} {\bibinfo {author} {\bibfnamefont {L.}~\bibnamefont
  {Amendola}}\ and\ \bibinfo {author} {\bibfnamefont {S.}~\bibnamefont
  {Tsujikawa}},\ }\href@noop {} {\emph {\bibinfo {title} {{Dark Energy, Theory
  and Observations}}}}\ (\bibinfo  {publisher} {Cambridge University Press},\
  \bibinfo {address} {Cambridge, England},\ \bibinfo {year} {2010})\BibitemShut
  {NoStop}%
\bibitem [{\citenamefont {Riess}(2019)}]{Riess:2019qba}%
  \BibitemOpen
  \bibfield  {author} {\bibinfo {author} {\bibfnamefont {A.~G.}\ \bibnamefont
  {Riess}},\ }\href {\doibase 10.1038/s42254-019-0137-0} {\bibfield  {journal}
  {\bibinfo  {journal} {Nature Rev. Phys.}\ }\textbf {\bibinfo {volume} {2}},\
  \bibinfo {pages} {10} (\bibinfo {year} {2019})},\ \Eprint
  {http://arxiv.org/abs/2001.03624} {arXiv:2001.03624 [astro-ph.CO]}
  \BibitemShut {NoStop}%
\bibitem [{\citenamefont {Di~Valentino}\ \emph {et~al.}(2021)\citenamefont
  {Di~Valentino}, \citenamefont {Mena}, \citenamefont {Pan}, \citenamefont
  {Visinelli}, \citenamefont {Yang}, \citenamefont {Melchiorri}, \citenamefont
  {Mota}, \citenamefont {Riess},\ and\ \citenamefont
  {Silk}}]{DiValentino:2021izs}%
  \BibitemOpen
  \bibfield  {author} {\bibinfo {author} {\bibfnamefont {E.}~\bibnamefont
  {Di~Valentino}}, \bibinfo {author} {\bibfnamefont {O.}~\bibnamefont {Mena}},
  \bibinfo {author} {\bibfnamefont {S.}~\bibnamefont {Pan}}, \bibinfo {author}
  {\bibfnamefont {L.}~\bibnamefont {Visinelli}}, \bibinfo {author}
  {\bibfnamefont {W.}~\bibnamefont {Yang}}, \bibinfo {author} {\bibfnamefont
  {A.}~\bibnamefont {Melchiorri}}, \bibinfo {author} {\bibfnamefont {D.~F.}\
  \bibnamefont {Mota}}, \bibinfo {author} {\bibfnamefont {A.~G.}\ \bibnamefont
  {Riess}}, \ and\ \bibinfo {author} {\bibfnamefont {J.}~\bibnamefont {Silk}},\
  }\href {\doibase 10.1088/1361-6382/ac086d} {\bibfield  {journal} {\bibinfo
  {journal} {Class. Quant. Grav.}\ }\textbf {\bibinfo {volume} {38}},\ \bibinfo
  {pages} {153001} (\bibinfo {year} {2021})},\ \Eprint
  {http://arxiv.org/abs/2103.01183} {arXiv:2103.01183 [astro-ph.CO]}
  \BibitemShut {NoStop}%
\bibitem [{\citenamefont {Brans}\ and\ \citenamefont
  {Dicke}(1961)}]{Brans:1961sx}%
  \BibitemOpen
  \bibfield  {author} {\bibinfo {author} {\bibfnamefont {C.}~\bibnamefont
  {Brans}}\ and\ \bibinfo {author} {\bibfnamefont {R.~H.}\ \bibnamefont
  {Dicke}},\ }\href {\doibase 10.1103/PhysRev.124.925} {\bibfield  {journal}
  {\bibinfo  {journal} {Phys. Rev.}\ }\textbf {\bibinfo {volume} {124}},\
  \bibinfo {pages} {925} (\bibinfo {year} {1961})}\BibitemShut {NoStop}%
\bibitem [{\citenamefont {Bergmann}(1968)}]{Bergmann:1968ve}%
  \BibitemOpen
  \bibfield  {author} {\bibinfo {author} {\bibfnamefont {P.~G.}\ \bibnamefont
  {Bergmann}},\ }\href {\doibase 10.1007/BF00668828} {\bibfield  {journal}
  {\bibinfo  {journal} {Int. J. Theor. Phys.}\ }\textbf {\bibinfo {volume}
  {1}},\ \bibinfo {pages} {25} (\bibinfo {year} {1968})}\BibitemShut {NoStop}%
\bibitem [{\citenamefont {Nordtvedt}(1968)}]{Nordtvedt:1968qs}%
  \BibitemOpen
  \bibfield  {author} {\bibinfo {author} {\bibfnamefont {K.}~\bibnamefont
  {Nordtvedt}},\ }\href {\doibase 10.1103/PhysRev.169.1017} {\bibfield
  {journal} {\bibinfo  {journal} {Phys. Rev.}\ }\textbf {\bibinfo {volume}
  {169}},\ \bibinfo {pages} {1017} (\bibinfo {year} {1968})}\BibitemShut
  {NoStop}%
\bibitem [{\citenamefont {Wagoner}(1970)}]{Wagoner:1970vr}%
  \BibitemOpen
  \bibfield  {author} {\bibinfo {author} {\bibfnamefont {R.~V.}\ \bibnamefont
  {Wagoner}},\ }\href {\doibase 10.1103/PhysRevD.1.3209} {\bibfield  {journal}
  {\bibinfo  {journal} {Phys. Rev. D}\ }\textbf {\bibinfo {volume} {1}},\
  \bibinfo {pages} {3209} (\bibinfo {year} {1970})}\BibitemShut {NoStop}%
\bibitem [{\citenamefont {Nordtvedt}(1970)}]{Nordtvedt:1970uv}%
  \BibitemOpen
  \bibfield  {author} {\bibinfo {author} {\bibfnamefont {K.}~\bibnamefont
  {Nordtvedt}, \bibfnamefont {Jr.}},\ }\href {\doibase 10.1086/150607}
  {\bibfield  {journal} {\bibinfo  {journal} {Astrophys. J.}\ }\textbf
  {\bibinfo {volume} {161}},\ \bibinfo {pages} {1059} (\bibinfo {year}
  {1970})}\BibitemShut {NoStop}%
\bibitem [{\citenamefont {Capozziello}\ \emph {et~al.}(2003)\citenamefont
  {Capozziello}, \citenamefont {Carloni},\ and\ \citenamefont
  {Troisi}}]{Capozziello:2003tk}%
  \BibitemOpen
  \bibfield  {author} {\bibinfo {author} {\bibfnamefont {S.}~\bibnamefont
  {Capozziello}}, \bibinfo {author} {\bibfnamefont {S.}~\bibnamefont
  {Carloni}}, \ and\ \bibinfo {author} {\bibfnamefont {A.}~\bibnamefont
  {Troisi}},\ }\href@noop {} {\bibfield  {journal} {\bibinfo  {journal} {Recent
  Res. Dev. Astron. Astrophys.}\ }\textbf {\bibinfo {volume} {1}},\ \bibinfo
  {pages} {625} (\bibinfo {year} {2003})},\ \Eprint
  {http://arxiv.org/abs/astro-ph/0303041} {arXiv:astro-ph/0303041} \BibitemShut
  {NoStop}%
\bibitem [{\citenamefont {Sotiriou}\ and\ \citenamefont
  {Faraoni}(2010)}]{Sotiriou:2008rp}%
  \BibitemOpen
  \bibfield  {author} {\bibinfo {author} {\bibfnamefont {T.~P.}\ \bibnamefont
  {Sotiriou}}\ and\ \bibinfo {author} {\bibfnamefont {V.}~\bibnamefont
  {Faraoni}},\ }\href {\doibase 10.1103/RevModPhys.82.451} {\bibfield
  {journal} {\bibinfo  {journal} {Rev. Mod. Phys.}\ }\textbf {\bibinfo {volume}
  {82}},\ \bibinfo {pages} {451} (\bibinfo {year} {2010})},\ \Eprint
  {http://arxiv.org/abs/0805.1726} {arXiv:0805.1726 [gr-qc]} \BibitemShut
  {NoStop}%
\bibitem [{\citenamefont {De~Felice}\ and\ \citenamefont
  {Tsujikawa}(2010)}]{DeFelice:2010aj}%
  \BibitemOpen
  \bibfield  {author} {\bibinfo {author} {\bibfnamefont {A.}~\bibnamefont
  {De~Felice}}\ and\ \bibinfo {author} {\bibfnamefont {S.}~\bibnamefont
  {Tsujikawa}},\ }\href {\doibase 10.12942/lrr-2010-3} {\bibfield  {journal}
  {\bibinfo  {journal} {Living Rev. Rel.}\ }\textbf {\bibinfo {volume} {13}},\
  \bibinfo {pages} {3} (\bibinfo {year} {2010})},\ \Eprint
  {http://arxiv.org/abs/1002.4928} {arXiv:1002.4928 [gr-qc]} \BibitemShut
  {NoStop}%
\bibitem [{\citenamefont {Nojiri}\ and\ \citenamefont
  {Odintsov}(2011)}]{Nojiri:2010wj}%
  \BibitemOpen
  \bibfield  {author} {\bibinfo {author} {\bibfnamefont {S.}~\bibnamefont
  {Nojiri}}\ and\ \bibinfo {author} {\bibfnamefont {S.~D.}\ \bibnamefont
  {Odintsov}},\ }\href {\doibase 10.1016/j.physrep.2011.04.001} {\bibfield
  {journal} {\bibinfo  {journal} {Phys. Rept.}\ }\textbf {\bibinfo {volume}
  {505}},\ \bibinfo {pages} {59} (\bibinfo {year} {2011})},\ \Eprint
  {http://arxiv.org/abs/1011.0544} {arXiv:1011.0544 [gr-qc]} \BibitemShut
  {NoStop}%
\bibitem [{\citenamefont {Horndeski}(1974)}]{Horndeski}%
  \BibitemOpen
  \bibfield  {author} {\bibinfo {author} {\bibfnamefont {G.~W.}\ \bibnamefont
  {Horndeski}},\ }\href {\doibase 10.1007/BF01807638} {\bibfield  {journal}
  {\bibinfo  {journal} {Int. J. Theor. Phys.}\ }\textbf {\bibinfo {volume}
  {10}},\ \bibinfo {pages} {363} (\bibinfo {year} {1974})}\BibitemShut
  {NoStop}%
\bibitem [{\citenamefont {Kobayashi}(2019)}]{Kobayashi:2019hrl}%
  \BibitemOpen
  \bibfield  {author} {\bibinfo {author} {\bibfnamefont {T.}~\bibnamefont
  {Kobayashi}},\ }\href {\doibase 10.1088/1361-6633/ab2429} {\bibfield
  {journal} {\bibinfo  {journal} {Rept. Prog. Phys.}\ }\textbf {\bibinfo
  {volume} {82}},\ \bibinfo {pages} {086901} (\bibinfo {year} {2019})},\
  \Eprint {http://arxiv.org/abs/1901.07183} {arXiv:1901.07183 [gr-qc]}
  \BibitemShut {NoStop}%
\bibitem [{\citenamefont {Langlois}(2019)}]{DHOSTreview1}%
  \BibitemOpen
  \bibfield  {author} {\bibinfo {author} {\bibfnamefont {D.}~\bibnamefont
  {Langlois}},\ }\href {\doibase 10.1142/S0218271819420069} {\bibfield
  {journal} {\bibinfo  {journal} {Int. J. Mod. Phys. D}\ }\textbf {\bibinfo
  {volume} {28}},\ \bibinfo {pages} {1942006} (\bibinfo {year} {2019})},\
  \Eprint {http://arxiv.org/abs/1811.06271} {arXiv:1811.06271 [gr-qc]}
  \BibitemShut {NoStop}%
\bibitem [{\citenamefont {Abbott}\ \emph
  {et~al.}(2017{\natexlab{a}})\citenamefont {Abbott} \emph
  {et~al.}}]{LIGOScientific:2017vwq}%
  \BibitemOpen
  \bibfield  {author} {\bibinfo {author} {\bibfnamefont {B.~P.}\ \bibnamefont
  {Abbott}} \emph {et~al.} (\bibinfo {collaboration} {LIGO Scientific,
  Virgo}),\ }\href {\doibase 10.1103/PhysRevLett.119.161101} {\bibfield
  {journal} {\bibinfo  {journal} {Phys. Rev. Lett.}\ }\textbf {\bibinfo
  {volume} {119}},\ \bibinfo {pages} {161101} (\bibinfo {year}
  {2017}{\natexlab{a}})},\ \Eprint {http://arxiv.org/abs/1710.05832}
  {arXiv:1710.05832 [gr-qc]} \BibitemShut {NoStop}%
\bibitem [{\citenamefont {Abbott}\ \emph
  {et~al.}(2017{\natexlab{b}})\citenamefont {Abbott} \emph
  {et~al.}}]{LIGOScientific:2017zic}%
  \BibitemOpen
  \bibfield  {author} {\bibinfo {author} {\bibfnamefont {B.~P.}\ \bibnamefont
  {Abbott}} \emph {et~al.} (\bibinfo {collaboration} {LIGO Scientific, Virgo,
  Fermi-GBM, INTEGRAL}),\ }\href {\doibase 10.3847/2041-8213/aa920c} {\bibfield
   {journal} {\bibinfo  {journal} {Astrophys. J. Lett.}\ }\textbf {\bibinfo
  {volume} {848}},\ \bibinfo {pages} {L13} (\bibinfo {year}
  {2017}{\natexlab{b}})},\ \Eprint {http://arxiv.org/abs/1710.05834}
  {arXiv:1710.05834 [astro-ph.HE]} \BibitemShut {NoStop}%
\bibitem [{\citenamefont {Langlois}\ \emph {et~al.}(2018)\citenamefont
  {Langlois}, \citenamefont {Saito}, \citenamefont {Yamauchi},\ and\
  \citenamefont {Noui}}]{Langlois:2017dyl}%
  \BibitemOpen
  \bibfield  {author} {\bibinfo {author} {\bibfnamefont {D.}~\bibnamefont
  {Langlois}}, \bibinfo {author} {\bibfnamefont {R.}~\bibnamefont {Saito}},
  \bibinfo {author} {\bibfnamefont {D.}~\bibnamefont {Yamauchi}}, \ and\
  \bibinfo {author} {\bibfnamefont {K.}~\bibnamefont {Noui}},\ }\href {\doibase
  10.1103/PhysRevD.97.061501} {\bibfield  {journal} {\bibinfo  {journal} {Phys.
  Rev. D}\ }\textbf {\bibinfo {volume} {97}},\ \bibinfo {pages} {061501}
  (\bibinfo {year} {2018})},\ \Eprint {http://arxiv.org/abs/1711.07403}
  {arXiv:1711.07403 [gr-qc]} \BibitemShut {NoStop}%
\bibitem [{\citenamefont {Pimentel}(1989)}]{Pimentel:1989bm}%
  \BibitemOpen
  \bibfield  {author} {\bibinfo {author} {\bibfnamefont {L.~O.}\ \bibnamefont
  {Pimentel}},\ }\href {\doibase 10.1088/0264-9381/6/12/005} {\bibfield
  {journal} {\bibinfo  {journal} {Class. Quant. Grav.}\ }\textbf {\bibinfo
  {volume} {6}},\ \bibinfo {pages} {L263} (\bibinfo {year} {1989})}\BibitemShut
  {NoStop}%
\bibitem [{\citenamefont {Faraoni}\ and\ \citenamefont
  {Cot\'e}(2018)}]{Faraoni:2018qdr}%
  \BibitemOpen
  \bibfield  {author} {\bibinfo {author} {\bibfnamefont {V.}~\bibnamefont
  {Faraoni}}\ and\ \bibinfo {author} {\bibfnamefont {J.}~\bibnamefont
  {Cot\'e}},\ }\href {\doibase 10.1103/PhysRevD.98.084019} {\bibfield
  {journal} {\bibinfo  {journal} {Phys. Rev. D}\ }\textbf {\bibinfo {volume}
  {98}},\ \bibinfo {pages} {084019} (\bibinfo {year} {2018})},\ \Eprint
  {http://arxiv.org/abs/1808.02427} {arXiv:1808.02427 [gr-qc]} \BibitemShut
  {NoStop}%
\bibitem [{\citenamefont {Nucamendi}\ \emph {et~al.}(2020)\citenamefont
  {Nucamendi}, \citenamefont {De~Arcia}, \citenamefont {Gonzalez},
  \citenamefont {Horta-Rangel},\ and\ \citenamefont
  {Quiros}}]{Nucamendi:2019uen}%
  \BibitemOpen
  \bibfield  {author} {\bibinfo {author} {\bibfnamefont {U.}~\bibnamefont
  {Nucamendi}}, \bibinfo {author} {\bibfnamefont {R.}~\bibnamefont {De~Arcia}},
  \bibinfo {author} {\bibfnamefont {T.}~\bibnamefont {Gonzalez}}, \bibinfo
  {author} {\bibfnamefont {F.~A.}\ \bibnamefont {Horta-Rangel}}, \ and\
  \bibinfo {author} {\bibfnamefont {I.}~\bibnamefont {Quiros}},\ }\href
  {\doibase 10.1103/PhysRevD.102.084054} {\bibfield  {journal} {\bibinfo
  {journal} {Phys. Rev. D}\ }\textbf {\bibinfo {volume} {102}},\ \bibinfo
  {pages} {084054} (\bibinfo {year} {2020})},\ \Eprint
  {http://arxiv.org/abs/1910.13026} {arXiv:1910.13026 [gr-qc]} \BibitemShut
  {NoStop}%
\bibitem [{\citenamefont {Giusti}\ \emph {et~al.}(2022)\citenamefont {Giusti},
  \citenamefont {Zentarra}, \citenamefont {Heisenberg},\ and\ \citenamefont
  {Faraoni}}]{Giusti:2021sku}%
  \BibitemOpen
  \bibfield  {author} {\bibinfo {author} {\bibfnamefont {A.}~\bibnamefont
  {Giusti}}, \bibinfo {author} {\bibfnamefont {S.}~\bibnamefont {Zentarra}},
  \bibinfo {author} {\bibfnamefont {L.}~\bibnamefont {Heisenberg}}, \ and\
  \bibinfo {author} {\bibfnamefont {V.}~\bibnamefont {Faraoni}},\ }\href
  {\doibase 10.1103/PhysRevD.105.124011} {\bibfield  {journal} {\bibinfo
  {journal} {Phys. Rev. D}\ }\textbf {\bibinfo {volume} {105}},\ \bibinfo
  {pages} {124011} (\bibinfo {year} {2022})},\ \Eprint
  {http://arxiv.org/abs/2108.10706} {arXiv:2108.10706 [gr-qc]} \BibitemShut
  {NoStop}%
\bibitem [{\citenamefont {Eckart}(1940)}]{Eckart:1940te}%
  \BibitemOpen
  \bibfield  {author} {\bibinfo {author} {\bibfnamefont {C.}~\bibnamefont
  {Eckart}},\ }\href {\doibase 10.1103/PhysRev.58.919} {\bibfield  {journal}
  {\bibinfo  {journal} {Phys. Rev.}\ }\textbf {\bibinfo {volume} {58}},\
  \bibinfo {pages} {919} (\bibinfo {year} {1940})}\BibitemShut {NoStop}%
\bibitem [{\citenamefont {Faraoni}\ and\ \citenamefont
  {Giusti}(2021)}]{Faraoni:2021lfc}%
  \BibitemOpen
  \bibfield  {author} {\bibinfo {author} {\bibfnamefont {V.}~\bibnamefont
  {Faraoni}}\ and\ \bibinfo {author} {\bibfnamefont {A.}~\bibnamefont
  {Giusti}},\ }\href {\doibase 10.1103/PhysRevD.103.L121501} {\bibfield
  {journal} {\bibinfo  {journal} {Phys. Rev. D}\ }\textbf {\bibinfo {volume}
  {103}},\ \bibinfo {pages} {L121501} (\bibinfo {year} {2021})},\ \Eprint
  {http://arxiv.org/abs/2103.05389} {arXiv:2103.05389 [gr-qc]} \BibitemShut
  {NoStop}%
\bibitem [{\citenamefont {Faraoni}\ \emph
  {et~al.}(2021{\natexlab{a}})\citenamefont {Faraoni}, \citenamefont {Giusti},\
  and\ \citenamefont {Mentrelli}}]{Faraoni:2021jri}%
  \BibitemOpen
  \bibfield  {author} {\bibinfo {author} {\bibfnamefont {V.}~\bibnamefont
  {Faraoni}}, \bibinfo {author} {\bibfnamefont {A.}~\bibnamefont {Giusti}}, \
  and\ \bibinfo {author} {\bibfnamefont {A.}~\bibnamefont {Mentrelli}},\ }\href
  {\doibase 10.1103/PhysRevD.104.124031} {\bibfield  {journal} {\bibinfo
  {journal} {Phys. Rev. D}\ }\textbf {\bibinfo {volume} {104}},\ \bibinfo
  {pages} {124031} (\bibinfo {year} {2021}{\natexlab{a}})},\ \Eprint
  {http://arxiv.org/abs/2110.02368} {arXiv:2110.02368 [gr-qc]} \BibitemShut
  {NoStop}%
\bibitem [{\citenamefont {Jacobson}(1995)}]{Jacobson:1995ab}%
  \BibitemOpen
  \bibfield  {author} {\bibinfo {author} {\bibfnamefont {T.}~\bibnamefont
  {Jacobson}},\ }\href {\doibase 10.1103/PhysRevLett.75.1260} {\bibfield
  {journal} {\bibinfo  {journal} {Phys. Rev. Lett.}\ }\textbf {\bibinfo
  {volume} {75}},\ \bibinfo {pages} {1260} (\bibinfo {year} {1995})},\ \Eprint
  {http://arxiv.org/abs/gr-qc/9504004} {arXiv:gr-qc/9504004} \BibitemShut
  {NoStop}%
\bibitem [{\citenamefont {Eling}\ \emph {et~al.}(2006)\citenamefont {Eling},
  \citenamefont {Guedens},\ and\ \citenamefont {Jacobson}}]{Eling:2006aw}%
  \BibitemOpen
  \bibfield  {author} {\bibinfo {author} {\bibfnamefont {C.}~\bibnamefont
  {Eling}}, \bibinfo {author} {\bibfnamefont {R.}~\bibnamefont {Guedens}}, \
  and\ \bibinfo {author} {\bibfnamefont {T.}~\bibnamefont {Jacobson}},\ }\href
  {\doibase 10.1103/PhysRevLett.96.121301} {\bibfield  {journal} {\bibinfo
  {journal} {Phys. Rev. Lett.}\ }\textbf {\bibinfo {volume} {96}},\ \bibinfo
  {pages} {121301} (\bibinfo {year} {2006})},\ \Eprint
  {http://arxiv.org/abs/gr-qc/0602001} {arXiv:gr-qc/0602001} \BibitemShut
  {NoStop}%
\bibitem [{\citenamefont {Giardino}\ \emph {et~al.}(2022)\citenamefont
  {Giardino}, \citenamefont {Faraoni},\ and\ \citenamefont
  {Giusti}}]{Giardino:2022sdv}%
  \BibitemOpen
  \bibfield  {author} {\bibinfo {author} {\bibfnamefont {S.}~\bibnamefont
  {Giardino}}, \bibinfo {author} {\bibfnamefont {V.}~\bibnamefont {Faraoni}}, \
  and\ \bibinfo {author} {\bibfnamefont {A.}~\bibnamefont {Giusti}},\ }\href
  {\doibase 10.1088/1475-7516/2022/04/053} {\bibfield  {journal} {\bibinfo
  {journal} {JCAP}\ }\textbf {\bibinfo {volume} {04}},\ \bibinfo {pages} {053}
  (\bibinfo {year} {2022})},\ \Eprint {http://arxiv.org/abs/2202.07393}
  {arXiv:2202.07393 [gr-qc]} \BibitemShut {NoStop}%
\bibitem [{\citenamefont {Faraoni}\ and\ \citenamefont
  {Fran\c{c}onnet}(2022)}]{Faraoni:2022jyd}%
  \BibitemOpen
  \bibfield  {author} {\bibinfo {author} {\bibfnamefont {V.}~\bibnamefont
  {Faraoni}}\ and\ \bibinfo {author} {\bibfnamefont {T.~B.}\ \bibnamefont
  {Fran\c{c}onnet}},\ }\href {\doibase 10.1103/PhysRevD.105.104006} {\bibfield
  {journal} {\bibinfo  {journal} {Phys. Rev. D}\ }\textbf {\bibinfo {volume}
  {105}},\ \bibinfo {pages} {104006} (\bibinfo {year} {2022})},\ \Eprint
  {http://arxiv.org/abs/2203.14934} {arXiv:2203.14934 [gr-qc]} \BibitemShut
  {NoStop}%
\bibitem [{\citenamefont {Faraoni}\ \emph
  {et~al.}(2022{\natexlab{a}})\citenamefont {Faraoni}, \citenamefont {Giusti},
  \citenamefont {Jose},\ and\ \citenamefont {Giardino}}]{Faraoni:2022doe}%
  \BibitemOpen
  \bibfield  {author} {\bibinfo {author} {\bibfnamefont {V.}~\bibnamefont
  {Faraoni}}, \bibinfo {author} {\bibfnamefont {A.}~\bibnamefont {Giusti}},
  \bibinfo {author} {\bibfnamefont {S.}~\bibnamefont {Jose}}, \ and\ \bibinfo
  {author} {\bibfnamefont {S.}~\bibnamefont {Giardino}},\ }\href {\doibase
  10.1103/PhysRevD.106.024049} {\bibfield  {journal} {\bibinfo  {journal}
  {Phys. Rev. D}\ }\textbf {\bibinfo {volume} {106}},\ \bibinfo {pages}
  {024049} (\bibinfo {year} {2022}{\natexlab{a}})},\ \Eprint
  {http://arxiv.org/abs/2206.02046} {arXiv:2206.02046 [gr-qc]} \BibitemShut
  {NoStop}%
\bibitem [{\citenamefont {Faraoni}\ \emph
  {et~al.}(2022{\natexlab{b}})\citenamefont {Faraoni}, \citenamefont
  {Giardino}, \citenamefont {Giusti},\ and\ \citenamefont
  {Vanderwee}}]{Faraoni:2022gry}%
  \BibitemOpen
  \bibfield  {author} {\bibinfo {author} {\bibfnamefont {V.}~\bibnamefont
  {Faraoni}}, \bibinfo {author} {\bibfnamefont {S.}~\bibnamefont {Giardino}},
  \bibinfo {author} {\bibfnamefont {A.}~\bibnamefont {Giusti}}, \ and\ \bibinfo
  {author} {\bibfnamefont {R.}~\bibnamefont {Vanderwee}},\ }\href@noop {} {\
  (\bibinfo {year} {2022}{\natexlab{b}})},\ \Eprint
  {http://arxiv.org/abs/2208.04051} {arXiv:2208.04051 [gr-qc]} \BibitemShut
  {NoStop}%
\bibitem [{\citenamefont {Wald}(1984)}]{Waldbook}%
  \BibitemOpen
  \bibfield  {author} {\bibinfo {author} {\bibfnamefont {R.~M.}\ \bibnamefont
  {Wald}},\ }\href {\doibase 10.7208/chicago/9780226870373.001.0001} {\emph
  {\bibinfo {title} {{General Relativity}}}}\ (\bibinfo  {publisher} {Chicago
  Univ. Press},\ \bibinfo {address} {Chicago, USA},\ \bibinfo {year}
  {1984})\BibitemShut {NoStop}%
\bibitem [{\citenamefont {Faraoni}(2004)}]{mybook1}%
  \BibitemOpen
  \bibfield  {author} {\bibinfo {author} {\bibfnamefont {V.}~\bibnamefont
  {Faraoni}},\ }\href {\doibase 10.1007/978-1-4020-1989-0} {\emph {\bibinfo
  {title} {{Cosmology in Scalar-Tensor Gravity}}}}\ (\bibinfo  {publisher}
  {Kluwer Academic, Dordrecht},\ \bibinfo {year} {2004})\BibitemShut {NoStop}%
\bibitem [{\citenamefont {Faraoni}\ \emph
  {et~al.}(2021{\natexlab{b}})\citenamefont {Faraoni}, \citenamefont {Giusti},\
  and\ \citenamefont {Fahim}}]{Faraoni:2021nhi}%
  \BibitemOpen
  \bibfield  {author} {\bibinfo {author} {\bibfnamefont {V.}~\bibnamefont
  {Faraoni}}, \bibinfo {author} {\bibfnamefont {A.}~\bibnamefont {Giusti}}, \
  and\ \bibinfo {author} {\bibfnamefont {B.~H.}\ \bibnamefont {Fahim}},\ }\href
  {\doibase 10.1016/j.physrep.2021.04.003} {\bibfield  {journal} {\bibinfo
  {journal} {Phys. Rept.}\ }\textbf {\bibinfo {volume} {925}},\ \bibinfo
  {pages} {1} (\bibinfo {year} {2021}{\natexlab{b}})},\ \Eprint
  {http://arxiv.org/abs/2101.00266} {arXiv:2101.00266 [gr-qc]} \BibitemShut
  {NoStop}%
\bibitem [{\citenamefont {Ben~Achour}\ \emph
  {et~al.}(2020{\natexlab{a}})\citenamefont {Ben~Achour}, \citenamefont {Liu},\
  and\ \citenamefont {Mukohyama}}]{BenAchour:2020wiw}%
  \BibitemOpen
  \bibfield  {author} {\bibinfo {author} {\bibfnamefont {J.}~\bibnamefont
  {Ben~Achour}}, \bibinfo {author} {\bibfnamefont {H.}~\bibnamefont {Liu}}, \
  and\ \bibinfo {author} {\bibfnamefont {S.}~\bibnamefont {Mukohyama}},\ }\href
  {\doibase 10.1088/1475-7516/2020/02/023} {\bibfield  {journal} {\bibinfo
  {journal} {JCAP}\ }\textbf {\bibinfo {volume} {02}},\ \bibinfo {pages} {023}
  (\bibinfo {year} {2020}{\natexlab{a}})},\ \Eprint
  {http://arxiv.org/abs/1910.11017} {arXiv:1910.11017 [gr-qc]} \BibitemShut
  {NoStop}%
\bibitem [{\citenamefont {Faraoni}\ and\ \citenamefont
  {Leblanc}(2021)}]{Faraoni:2021gdl}%
  \BibitemOpen
  \bibfield  {author} {\bibinfo {author} {\bibfnamefont {V.}~\bibnamefont
  {Faraoni}}\ and\ \bibinfo {author} {\bibfnamefont {A.}~\bibnamefont
  {Leblanc}},\ }\href {\doibase 10.1088/1475-7516/2021/08/037} {\bibfield
  {journal} {\bibinfo  {journal} {JCAP}\ }\textbf {\bibinfo {volume} {08}},\
  \bibinfo {pages} {037} (\bibinfo {year} {2021})},\ \Eprint
  {http://arxiv.org/abs/2107.03456} {arXiv:2107.03456 [gr-qc]} \BibitemShut
  {NoStop}%
\bibitem [{\citenamefont {Achour}\ \emph {et~al.}(2021)\citenamefont {Achour},
  \citenamefont {De~Felice}, \citenamefont {Gorji}, \citenamefont {Mukohyama},\
  and\ \citenamefont {Pookkillath}}]{Achour:2021pla}%
  \BibitemOpen
  \bibfield  {author} {\bibinfo {author} {\bibfnamefont {J.~B.}\ \bibnamefont
  {Achour}}, \bibinfo {author} {\bibfnamefont {A.}~\bibnamefont {De~Felice}},
  \bibinfo {author} {\bibfnamefont {M.~A.}\ \bibnamefont {Gorji}}, \bibinfo
  {author} {\bibfnamefont {S.}~\bibnamefont {Mukohyama}}, \ and\ \bibinfo
  {author} {\bibfnamefont {M.~C.}\ \bibnamefont {Pookkillath}},\ }\href
  {\doibase 10.1088/1475-7516/2021/10/067} {\bibfield  {journal} {\bibinfo
  {journal} {JCAP}\ }\textbf {\bibinfo {volume} {10}},\ \bibinfo {pages} {067}
  (\bibinfo {year} {2021})},\ \Eprint {http://arxiv.org/abs/2107.02386}
  {arXiv:2107.02386 [gr-qc]} \BibitemShut {NoStop}%
\bibitem [{\citenamefont {Babichev}\ and\ \citenamefont
  {Esposito-Far\`ese}(2013)}]{Babichev:2012re}%
  \BibitemOpen
  \bibfield  {author} {\bibinfo {author} {\bibfnamefont {E.}~\bibnamefont
  {Babichev}}\ and\ \bibinfo {author} {\bibfnamefont {G.}~\bibnamefont
  {Esposito-Far\`ese}},\ }\href {\doibase 10.1103/PhysRevD.87.044032}
  {\bibfield  {journal} {\bibinfo  {journal} {Phys. Rev. D}\ }\textbf {\bibinfo
  {volume} {87}},\ \bibinfo {pages} {044032} (\bibinfo {year} {2013})},\
  \Eprint {http://arxiv.org/abs/1212.1394} {arXiv:1212.1394 [gr-qc]}
  \BibitemShut {NoStop}%
\bibitem [{\citenamefont {Anabalon}\ \emph {et~al.}(2014)\citenamefont
  {Anabalon}, \citenamefont {Cisterna},\ and\ \citenamefont
  {Oliva}}]{Anabalon:2013oea}%
  \BibitemOpen
  \bibfield  {author} {\bibinfo {author} {\bibfnamefont {A.}~\bibnamefont
  {Anabalon}}, \bibinfo {author} {\bibfnamefont {A.}~\bibnamefont {Cisterna}},
  \ and\ \bibinfo {author} {\bibfnamefont {J.}~\bibnamefont {Oliva}},\ }\href
  {\doibase 10.1103/PhysRevD.89.084050} {\bibfield  {journal} {\bibinfo
  {journal} {Phys. Rev. D}\ }\textbf {\bibinfo {volume} {89}},\ \bibinfo
  {pages} {084050} (\bibinfo {year} {2014})},\ \Eprint
  {http://arxiv.org/abs/1312.3597} {arXiv:1312.3597 [gr-qc]} \BibitemShut
  {NoStop}%
\bibitem [{\citenamefont {Babichev}\ and\ \citenamefont
  {Charmousis}(2014)}]{Babichev:2013cya}%
  \BibitemOpen
  \bibfield  {author} {\bibinfo {author} {\bibfnamefont {E.}~\bibnamefont
  {Babichev}}\ and\ \bibinfo {author} {\bibfnamefont {C.}~\bibnamefont
  {Charmousis}},\ }\href {\doibase 10.1007/JHEP08(2014)106} {\bibfield
  {journal} {\bibinfo  {journal} {JHEP}\ }\textbf {\bibinfo {volume} {08}},\
  \bibinfo {pages} {106} (\bibinfo {year} {2014})},\ \Eprint
  {http://arxiv.org/abs/1312.3204} {arXiv:1312.3204 [gr-qc]} \BibitemShut
  {NoStop}%
\bibitem [{\citenamefont {Charmousis}\ \emph {et~al.}(2014)\citenamefont
  {Charmousis}, \citenamefont {Kolyvaris}, \citenamefont {Papantonopoulos},\
  and\ \citenamefont {Tsoukalas}}]{Charmousis:2014zaa}%
  \BibitemOpen
  \bibfield  {author} {\bibinfo {author} {\bibfnamefont {C.}~\bibnamefont
  {Charmousis}}, \bibinfo {author} {\bibfnamefont {T.}~\bibnamefont
  {Kolyvaris}}, \bibinfo {author} {\bibfnamefont {E.}~\bibnamefont
  {Papantonopoulos}}, \ and\ \bibinfo {author} {\bibfnamefont {M.}~\bibnamefont
  {Tsoukalas}},\ }\href {\doibase 10.1007/JHEP07(2014)085} {\bibfield
  {journal} {\bibinfo  {journal} {JHEP}\ }\textbf {\bibinfo {volume} {07}},\
  \bibinfo {pages} {085} (\bibinfo {year} {2014})},\ \Eprint
  {http://arxiv.org/abs/1404.1024} {arXiv:1404.1024 [gr-qc]} \BibitemShut
  {NoStop}%
\bibitem [{\citenamefont {Kobayashi}\ and\ \citenamefont
  {Tanahashi}(2014)}]{Kobayashi:2014eva}%
  \BibitemOpen
  \bibfield  {author} {\bibinfo {author} {\bibfnamefont {T.}~\bibnamefont
  {Kobayashi}}\ and\ \bibinfo {author} {\bibfnamefont {N.}~\bibnamefont
  {Tanahashi}},\ }\href {\doibase 10.1093/ptep/ptu096} {\bibfield  {journal}
  {\bibinfo  {journal} {PTEP}\ }\textbf {\bibinfo {volume} {2014}},\ \bibinfo
  {pages} {073E02} (\bibinfo {year} {2014})},\ \Eprint
  {http://arxiv.org/abs/1403.4364} {arXiv:1403.4364 [gr-qc]} \BibitemShut
  {NoStop}%
\bibitem [{\citenamefont {Babichev}\ and\ \citenamefont
  {Esposito-Farese}(2017)}]{Babichev:2016kdt}%
  \BibitemOpen
  \bibfield  {author} {\bibinfo {author} {\bibfnamefont {E.}~\bibnamefont
  {Babichev}}\ and\ \bibinfo {author} {\bibfnamefont {G.}~\bibnamefont
  {Esposito-Farese}},\ }\href {\doibase 10.1103/PhysRevD.95.024020} {\bibfield
  {journal} {\bibinfo  {journal} {Phys. Rev. D}\ }\textbf {\bibinfo {volume}
  {95}},\ \bibinfo {pages} {024020} (\bibinfo {year} {2017})},\ \Eprint
  {http://arxiv.org/abs/1609.09798} {arXiv:1609.09798 [gr-qc]} \BibitemShut
  {NoStop}%
\bibitem [{\citenamefont {Motohashi}\ and\ \citenamefont
  {Minamitsuji}(2018)}]{Motohashi:2018wdq}%
  \BibitemOpen
  \bibfield  {author} {\bibinfo {author} {\bibfnamefont {H.}~\bibnamefont
  {Motohashi}}\ and\ \bibinfo {author} {\bibfnamefont {M.}~\bibnamefont
  {Minamitsuji}},\ }\href {\doibase 10.1016/j.physletb.2018.04.041} {\bibfield
  {journal} {\bibinfo  {journal} {Phys. Lett. B}\ }\textbf {\bibinfo {volume}
  {781}},\ \bibinfo {pages} {728} (\bibinfo {year} {2018})},\ \Eprint
  {http://arxiv.org/abs/1804.01731} {arXiv:1804.01731 [gr-qc]} \BibitemShut
  {NoStop}%
\bibitem [{\citenamefont {Babichev}\ \emph {et~al.}(2017)\citenamefont
  {Babichev}, \citenamefont {Charmousis},\ and\ \citenamefont
  {Leh\'ebel}}]{Babichev:2017guv}%
  \BibitemOpen
  \bibfield  {author} {\bibinfo {author} {\bibfnamefont {E.}~\bibnamefont
  {Babichev}}, \bibinfo {author} {\bibfnamefont {C.}~\bibnamefont
  {Charmousis}}, \ and\ \bibinfo {author} {\bibfnamefont {A.}~\bibnamefont
  {Leh\'ebel}},\ }\href {\doibase 10.1088/1475-7516/2017/04/027} {\bibfield
  {journal} {\bibinfo  {journal} {JCAP}\ }\textbf {\bibinfo {volume} {04}},\
  \bibinfo {pages} {027} (\bibinfo {year} {2017})},\ \Eprint
  {http://arxiv.org/abs/1702.01938} {arXiv:1702.01938 [gr-qc]} \BibitemShut
  {NoStop}%
\bibitem [{\citenamefont {Anson}\ \emph {et~al.}(2021)\citenamefont {Anson},
  \citenamefont {Babichev}, \citenamefont {Charmousis},\ and\ \citenamefont
  {Hassaine}}]{Anson:2020trg}%
  \BibitemOpen
  \bibfield  {author} {\bibinfo {author} {\bibfnamefont {T.}~\bibnamefont
  {Anson}}, \bibinfo {author} {\bibfnamefont {E.}~\bibnamefont {Babichev}},
  \bibinfo {author} {\bibfnamefont {C.}~\bibnamefont {Charmousis}}, \ and\
  \bibinfo {author} {\bibfnamefont {M.}~\bibnamefont {Hassaine}},\ }\href
  {\doibase 10.1007/JHEP01(2021)018} {\bibfield  {journal} {\bibinfo  {journal}
  {JHEP}\ }\textbf {\bibinfo {volume} {01}},\ \bibinfo {pages} {018} (\bibinfo
  {year} {2021})},\ \Eprint {http://arxiv.org/abs/2006.06461} {arXiv:2006.06461
  [gr-qc]} \BibitemShut {NoStop}%
\bibitem [{\citenamefont {Ben~Achour}\ \emph
  {et~al.}(2020{\natexlab{b}})\citenamefont {Ben~Achour}, \citenamefont {Liu},
  \citenamefont {Motohashi}, \citenamefont {Mukohyama},\ and\ \citenamefont
  {Noui}}]{BenAchour:2020fgy}%
  \BibitemOpen
  \bibfield  {author} {\bibinfo {author} {\bibfnamefont {J.}~\bibnamefont
  {Ben~Achour}}, \bibinfo {author} {\bibfnamefont {H.}~\bibnamefont {Liu}},
  \bibinfo {author} {\bibfnamefont {H.}~\bibnamefont {Motohashi}}, \bibinfo
  {author} {\bibfnamefont {S.}~\bibnamefont {Mukohyama}}, \ and\ \bibinfo
  {author} {\bibfnamefont {K.}~\bibnamefont {Noui}},\ }\href {\doibase
  10.1088/1475-7516/2020/11/001} {\bibfield  {journal} {\bibinfo  {journal}
  {JCAP}\ }\textbf {\bibinfo {volume} {11}},\ \bibinfo {pages} {001} (\bibinfo
  {year} {2020}{\natexlab{b}})},\ \Eprint {http://arxiv.org/abs/2006.07245}
  {arXiv:2006.07245 [gr-qc]} \BibitemShut {NoStop}%
\bibitem [{\citenamefont {Chatzifotis}\ \emph {et~al.}(2022)\citenamefont
  {Chatzifotis}, \citenamefont {Papantonopoulos},\ and\ \citenamefont
  {Vlachos}}]{Chatzifotis:2021hpg}%
  \BibitemOpen
  \bibfield  {author} {\bibinfo {author} {\bibfnamefont {N.}~\bibnamefont
  {Chatzifotis}}, \bibinfo {author} {\bibfnamefont {E.}~\bibnamefont
  {Papantonopoulos}}, \ and\ \bibinfo {author} {\bibfnamefont {C.}~\bibnamefont
  {Vlachos}},\ }\href {\doibase 10.1103/PhysRevD.105.064025} {\bibfield
  {journal} {\bibinfo  {journal} {Phys. Rev. D}\ }\textbf {\bibinfo {volume}
  {105}},\ \bibinfo {pages} {064025} (\bibinfo {year} {2022})},\ \Eprint
  {http://arxiv.org/abs/2111.08773} {arXiv:2111.08773 [gr-qc]} \BibitemShut
  {NoStop}%
\bibitem [{\citenamefont {Saadati}\ \emph {et~al.}(2022)\citenamefont
  {Saadati}, \citenamefont {Giusti}, \citenamefont {Faraoni},\ and\
  \citenamefont {Shojai}}]{Saadati:2022jdc}%
  \BibitemOpen
  \bibfield  {author} {\bibinfo {author} {\bibfnamefont {R.}~\bibnamefont
  {Saadati}}, \bibinfo {author} {\bibfnamefont {A.}~\bibnamefont {Giusti}},
  \bibinfo {author} {\bibfnamefont {V.}~\bibnamefont {Faraoni}}, \ and\
  \bibinfo {author} {\bibfnamefont {F.}~\bibnamefont {Shojai}},\ }\href
  {\doibase 10.1088/1475-7516/2022/09/067} {\bibfield  {journal} {\bibinfo
  {journal} {JCAP}\ }\textbf {\bibinfo {volume} {09}},\ \bibinfo {pages} {067}
  (\bibinfo {year} {2022})},\ \Eprint {http://arxiv.org/abs/2207.07060}
  {arXiv:2207.07060 [gr-qc]} \BibitemShut {NoStop}%
\bibitem [{\citenamefont {Miranda}\ \emph {et~al.}(2022)\citenamefont
  {Miranda}, \citenamefont {Vernieri}, \citenamefont {Capozziello},\ and\
  \citenamefont {Faraoni}}]{Miranda:2022wkz}%
  \BibitemOpen
  \bibfield  {author} {\bibinfo {author} {\bibfnamefont {M.}~\bibnamefont
  {Miranda}}, \bibinfo {author} {\bibfnamefont {D.}~\bibnamefont {Vernieri}},
  \bibinfo {author} {\bibfnamefont {S.}~\bibnamefont {Capozziello}}, \ and\
  \bibinfo {author} {\bibfnamefont {V.}~\bibnamefont {Faraoni}},\ }\href@noop
  {} {\  (\bibinfo {year} {2022})},\ \Eprint {http://arxiv.org/abs/2209.02727}
  {arXiv:2209.02727 [gr-qc]} \BibitemShut {NoStop}%
\end{thebibliography}%

\end{document}